\documentclass[%
reprint, prd
]{revtex4-2}

\usepackage{graphicx}
\usepackage{dcolumn}
\usepackage{bm}
\usepackage{tikz}
\usepackage{amsmath}
\usepackage{amssymb}

\begin{document}

\preprint{APS/123-QED}

\title{Partial Derivatives on Causal Sets}

\author{Samuel Shuman}%
\email{shumans@oregonstate.edu}
\affiliation{Department of Physics, Oregon State University.}

\date{\today}

\begin{abstract}
    We will discuss two approaches to estimating partial derivatives and the metric components; one utilizing past work describing a causal set $\Box$ operator, and one using a construction from linear algebra called the Moore-Penrose inverse. After running numerical tests on a causal diamond in $\mathbb{M}^2$, we find that the approach using the Moore-Penrose inverse is significantly more accurate. Despite the large variances in the method using the $\Box$ operator, there is reason to believe both approaches should become more accurate at higher densities.
\end{abstract}

\maketitle


\section{\label{sec:intro} Introduction}
Causal Set Theory (CST) is a discrete theory of gravity that distinguishes itself from other similar theories by its minimal structure. When doing calculations on a causal set, the only tools provided are the causal order, $\prec$, and the average density, $\rho$. 

This is believed to be enough to recover the geometry of a Lorentzian manifold due to a series of papers starting with \cite{HawkingTopo} and \cite{MalamentTopo} in the late 1970s. These papers showed that the causal structure of a Lorentzian manifold is enough to recover the conformal geometry. Therefore, a discrete representation of spacetime should only need some representation of the causal structure and the volume to recover geometric properties \cite{BLSM}. 

For the ordering relation, $\prec$, to act as the causal structure, it must satisfy three properties.

\begin{enumerate}
    \item Transitivity: $x \prec y, y \prec z \Rightarrow x \prec z$
    \item Antisymmetry: $x \prec y \Rightarrow y \not\prec x$
    \item Local-Finiteness: $[x, y] = \{z | x \prec z \prec y\}$ is finite for all $x,y$
\end{enumerate}

Transitivity and antisymmetry ensure that $\prec$ acts like an ordering relation, while local-finiteness is necessary to ensure that volume information can be recovered from the discrete nature of spacetime. 

The relationship between volume and discreteness is encoded by the density, $\rho$. The volume of a region in a causal set is simply the number of events in the region divided by $\rho$. 

The simplest approach to modeling a discrete spacetime would be for events to be part of a regular lattice. Unfortunately, this approach does not work since the lattice vectors would break Lorentz invariance. Instead, the events in the causal set should be thought of as randomly selected from the background spacetime. This connection can also be used to generate causal sets by a process called sprinkling. The process starts by selecting a background manifold that we want to model with a causal set. Then we randomly select the events to include with a Poisson process at a density $\rho$. Finally, the causal order is inherited from the background manifold. 

There has been considerable research into how different geometric properties can be estimated on causal sets using these structures. For example, proper times \cite{CausalDiamond}, proper distances \cite{Overlap}, and curvature \cite{CausalDiamond, BoxOtherDims} can all be estimated using the causal order and the density. Further discussion of these estimators can be found in \cite{Surya}. In general relativity, however, these same properties are calculated using the tangent spaces. While we will not give an in depth review of tangent spaces here, it suffices to know that partial derivatives and the metric are key aspects of this formulation.

In this paper, we will develop two methods for calculating partial derivatives and metric components on causal sets. We will then test the results numerically to determine the efficacy of these approaches.

\subsection{Functions and Operators on Causal Sets \label{sec:functionsAndOps}}
Consider a finite region, $A$, of a causal set. On such a region, we can label the events with positive integer indices:
\[A = \{e_1, e_2, \dots, e_n\}\]
$n$ is the total number of events in the region, which is guaranteed to be finite since causal sets are locally-finite.

Now let us consider how to represent real-valued functions on this region. A function $f : A \to \mathbb{R}$, is uniquely defined by $n$ real numbers, $f_i = f(e_i)$. Thus we can represent a real-valued function on a finite region by a vector in $\mathbb{R}^n$. 

We will also need to represent differential operators on finite regions. A differential operator is a linear map that takes an input function and outputs a potentially different function that represents some sort of derivative of the input.

Since real-valued functions are represented by vectors in $\mathbb{R}^n$, we must represent differential operators by linear maps $\mathbb{R}^n \to \mathbb{R}^n$. This means we can represent differential operators with $n \times n$ matrices.

\subsubsection{The d'Alambertian}
In General Relativity (GR), the d'Alambertian operator is defined as $\Box = g^{\alpha\beta}\nabla_\alpha\nabla_\beta$. This operator has received particular interest in causal set research because it is a Lorentz invariant derivative, so it can be reasonably estimated from the causal structure. Consider the structure of a $\Box$ matrix operator acting on a scalar function $f$. 
\begin{equation}
    (B f)_i = \sum_j B_{ij} f_j
\end{equation}

As described in \cite{GeneralizedBox}, there are a few properties that the causal set $\Box$ operator is assumed to have. The first is retardedness, which means $(B f)_i$ should only depend on the values of $f$ at events preceding $e_i$. The next assumed property is ``label invariance", meaning the matrix is unchanged if you relabel the events in your causal set, which the authors call a discrete equivalent of general covariance. The last assumed property is called ``neighborly democracy", which requires the matrix components $B_{ij}$ to only depend on the number of events in the interval $[e_i, e_j]$. This ensures that the $\Box$ operator is Lorentz invariant. If we define $N(e_i, e_j) = |[e_i, e_j]|$, then we can express the matrix components as $B_{ij} = f(N(e_i, e_j))$.

\begin{figure}[h]
    \centering
    \includegraphics[width=8.6cm]{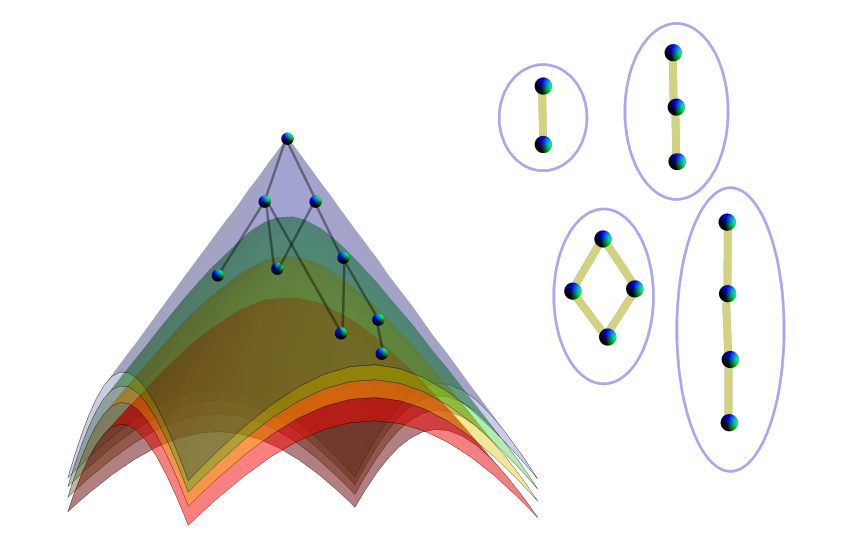}
    \caption{From \cite{Surya}, this shows the layered structure of the neighborhoods around a point. This also shows the types of 0,1, and 2 element intervals.}
    \label{fig:layers}
\end{figure}

The simplest candidate $\Box$ operator, first described in ~\cite{SorkinLocality}, defines 
\begin{equation}
    f(N(e_i, e_j)) = 
    \begin{cases}
        -\frac{1}{2}, \text{ if }e_i = e_j\\
        1, -2, 1, \text{ if } N = 0,1,2 \text{ respectively}\\
        0, \text{ otherwise}
    \end{cases}
\end{equation}
The matrix defined in this way was shown in \cite{SorkinLocality} to correctly estimate the d'Alambertian of simple test functions on average, in a sprinkling on $\mathbb{M}^2$. In other dimensions, other constants must be used in the definition of $\Box$ \cite{BoxOtherDims}.

There are two issues shared by these $\Box$ operators. First, despite having the correct average, the variation in $\Box f$ when compared across sprinklings does not go away at high $\rho$. In \cite{SorkinLocality}, it is argued that this variation is the result of the non-locality of causal set theory. The original $\Box$ operator defined above is affected by fluctuations at the discreteness scale. If we instead set the operator to depend on fluctuations at a set scale, those fluctuations will go to zero as the density goes to infinity. This is accomplished by modifying the $\Box$ operator in the following way.
\begin{equation}
    B_{ij} = 
    \begin{cases}
        -2\rho\epsilon, \text{ if } i = j\\
        4\rho\epsilon^2 f(N(e_i, e_j), \epsilon), \text{ if } e_i \prec e_j\\
    \end{cases}
\end{equation}
where we define
\begin{equation}
    f(n, \epsilon) = (1 - \epsilon)^n\bigg(1 - \frac{2\epsilon n}{1 - \epsilon} + \frac{\epsilon^2 n(n-1)}{2(1-\epsilon)^2}\bigg)
\end{equation}
The locality scale of this operator is $K = \epsilon \rho$, rather than $\rho$. Note that in the limit as $\epsilon = 1$, this recreates the original $\Box$ operator. \cite{SorkinLocality} claims the fluctuations in this operator now decrease like $\frac{1}{\sqrt{\rho}}$ when $K$ is held fixed. This derivation is extended to other dimensions in \cite{BoxOtherDims}. In this paper, we will fix $K = 200$. 

The second issue shared by these operators is that they do not average to the d'Alambertian in a curved spacetime. Instead, when applied to a scalar function $f$, the operators defined above average to $B f = \Box f - \frac{1}{2}R f$ \cite{BoxOtherDims}, where $R$ is the scalar curvature. To correct this, note that $\Box$ applied to a constant function should be zero. Thus we can assume that $B f = -\frac{1}{2} R$, when $f$ is the constant function $f = 1$. This allows us to isolate the $\Box$ operation from the action of the $B$ matrix for any scalar function. 
\begin{equation}
    \Box f = B f + \frac{1}{2} R f
\end{equation}

\subsubsection{Coordinate Functions}
In order to define partial derivatives on causal sets, we will need to represent coordinates. So what properties should our coordinates have? Coordinates on a causal set should be a set of functions $\{x^\alpha\}$ such that for all $e_i \neq e_j$ there is some $\beta$ such that $x^\beta(e_i) \neq x^\beta(e_j)$. Furthermore, these coordinate functions should be defined in such a way that, when averaged over sprinklings, the result is a smooth function on the manifold. The number of coordinate functions should match the dimension of the spacetime. 

One way to define these coordinate functions is found in \cite{JohnstonEmbedding}. The strategy the author uses is to embed the causal set in $\mathbb{M}^d$ for some dimension $d$. To do this, we must consider separation vectors between events in our causal set. The inner product of these separation vectors can be written in terms of the proper time estimators described in \cite{CausalDiamond}. For $e_i \prec e_j \prec e_k$
\begin{equation}
    \langle e_k - e_j, e_j - e_i \rangle = \frac{1}{2} \big(\tau_{ik}^2 - \tau_{ij}^2 - \tau_{jk}^2 \big)
\end{equation}
Now consider applying this strategy to an interval $[e_1, e_n]$. We will align a $t-$axis running between events $e_1$ and $e_n$. Let $T = \langle e_n - e_1, e_n - e_1 \rangle$. Then we define the $t-$coordinate on this interval as
\begin{equation}
    t(e_i) = \frac{1}{T}\langle e_i - e_1, e_n - e_1 \rangle
\end{equation}

\begin{figure}[h]
    \centering
    \includegraphics[width = 8.6cm]{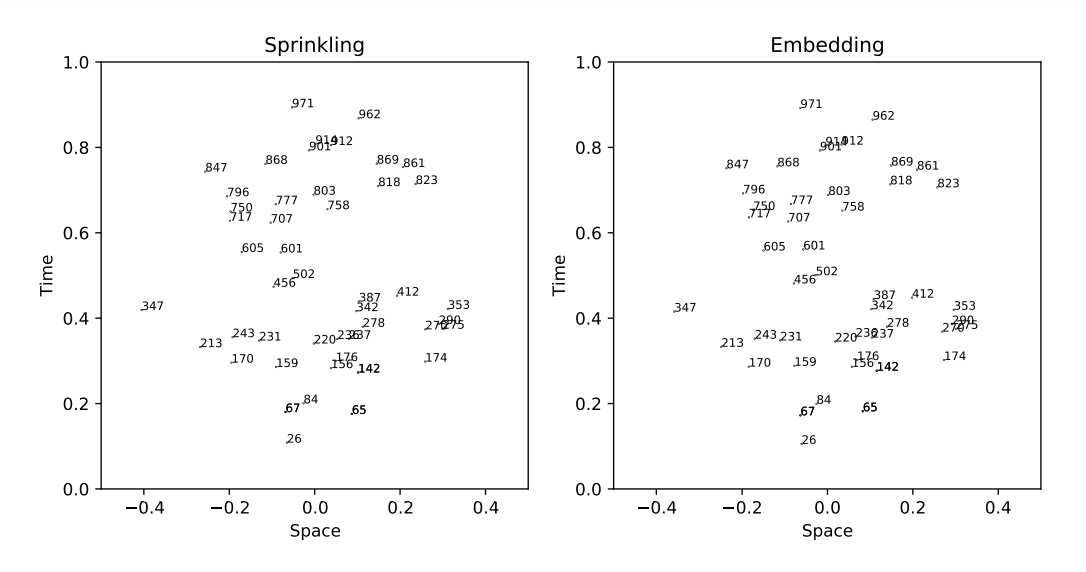}
    \caption{From ~\cite{JohnstonEmbedding}, this shows a random sample from an $n = 1000$ sprinkling and the resulting embedding, labeled by their indices.}
    \label{fig:embedding}
\end{figure}

It is more challenging to estimate spatial coordinates, but the author does so by defining a spatial inner product
\begin{equation}
    (e_k - e_j) \cdot (e_j - e_i) = (t_k - t_j)(t_j - t_i) - \langle e_k - e_j, e_j - e_i \rangle 
\end{equation}
In short, the spatial coordinates are then estimated by finding coordinate functions that match these spatial inner products. For a detailed discussion of how this works, see \cite{JohnstonEmbedding}. The result of this process is shown in Figure \ref{fig:embedding}. As we can see, these coordinate functions are an accurate estimate for the coordinates on the manifold. 

\subsection{The Moore-Penrose Inverse}
In many physical contexts, we encounter equations of the form $\vec{y} = A\vec{x}$, where $\vec{y}$ is a known vector, $A$ is a matrix, and we are looking for solutions $\vec{x}$. These types of equations exist in three categories. There are either (i) no solutions, (ii) exactly one solution, or (iii) infinitely many solutions. A common approach to solving such problems is using the matrix inverse $A^{-1}$, which allows us to find a solution $\vec{x} = A^{-1}\vec{y}$, however, this approach can only be applied when $A$ is a square matrix and the equation has exactly one solution, which is quite limiting. We will see that the Moore-Penrose inverse $A^+$ (sometimes called the generalized inverse) yields a good solution to any equation $\vec{y} = A \vec{x}$. 

To understand the construction of the Moore-Penrose inverse, we first must discuss singular value decomposition (SVD). SVD decomposes any matrix, $A$, into three parts: a diagonal matrix $\Sigma$, and two orthonormal matrices $U$ and $V$. The diagonal entries of $\Sigma$ are called the singular values. It is worth noting that every matrix $A$ can be decomposed in this way.
\begin{equation}
    A = U \Sigma V^T
\end{equation}

To construct the Moore-Penrose inverse of a matrix $A$, use SVD to get $A = U\Sigma V^T$. The Moore-Penrose inverse is given by $A^+ = V D U^T$, where $D$ is a diagonal matrix with entries,
\begin{equation}
    D_{ii} = 
    \begin{cases}
        0, \text{ if } \Sigma_{ii} = 0\\
        \frac{1}{\Sigma_{ii}}, \text{ otherwise}
    \end{cases}
\end{equation}

To understand why $A^+$ is a useful construction, let us return to the equation $\vec{y} = A \vec{x}$. The ``best" solution to this equation is given by $\vec{x} = A^+ \vec{y}$. The equation $\vec{y} = A \vec{x}$ has either zero, one, or infinitely many solutions. First, if the equation has exactly one solution, $A^+ \vec{y}$ will yield that solution. If, instead, the equation has infinitely many solutions, $A^+ \vec{y}$ will be the least-squares solution, in the sense that $||A^+ \vec{y}|| \leq ||\vec{x}||$ for any other solution $\vec{x}$. Lastly, and most importantly for our applications, if there is no solution to the equation, $A^+ \vec{y}$ will be the least-squares solution in the sense that $||\vec{y} - A \vec{x}||$ is minimized when $\vec{x} = A^+ \vec{y}$.

\section{\label{sec:boxDers}Partial Derivatives from the d'Alambertian}

\subsection{Theory \label{sec:boxDersTheory}}
In this section, we will discuss how you can combine coordinate estimators and the $\Box$ operator to get partial derivative operators. First, we will consider how the $\Box$ operator acts in the continuum. 

By definition, $\Box f = g^{\alpha\beta}\nabla_\alpha \nabla_\beta f$. $\nabla_\mu$ is the covariant derivative, which we have not discussed. The only property we will use is that for a scalar function $f$, $\nabla_\mu f = \partial_\mu f$. Consider how $\Box$ acts on a product of functions $f, h$ using the product rule.
\begin{align}
    &\Box (f h) = g^{\alpha\beta}\nabla_\alpha \nabla_\beta(f h) \nonumber \\
    &\Box (f h) = g^{\alpha \beta} \nabla_\alpha \left(f \nabla_\beta h + h\nabla_\beta f\right) \nonumber \\
    &\Box (f h) = g^{\alpha \beta} \big((\nabla_\alpha f) (\nabla_\beta h) + f \nabla_\alpha \nabla_\beta h \nonumber\\
    &\hspace{50pt} + h \nabla_\alpha \nabla_\beta f + (\nabla_\alpha h) (\nabla_\beta f) \big) \nonumber \\
    &\Box (f h) = f \Box h + h \Box f + 2g^{\alpha\beta}(\nabla_\alpha f) (\nabla_\beta h) \nonumber \\
    &g^{\alpha\beta} (\partial_\alpha f) (\partial_\beta h) = \frac{1}{2}\big(\Box (f h) - f \Box h - h \Box f \big) \label{doubleProduct}
\end{align}
Let us assume that our partial derivative operators and the inverse metric component functions on the causal set will satisfy equation \ref{doubleProduct} for any real functions $f$ and $h$. 

Now consider what happens if $f$ and $h$ are both coordinate functions.
\begin{equation}
    g^{\alpha\beta}(\partial_\alpha x^\mu)(\partial_\beta x^\nu) = \frac{1}{2} \big(\Box (x^\mu x^\nu) - x^\nu \Box x^\mu - x^\mu \Box x^\nu \big) \label{doubleProductCoords}
\end{equation}
If we assume that the partial derivative operators in the causal set keep another nice property from the continuum, namely that $\partial_\alpha x^\mu = \delta_\alpha^\mu$, then we can use equation \ref{doubleProductCoords} to calculate the inverse metric component functions. 
\begin{equation}
    g^{\mu\nu} = \frac{1}{2} \big(\Box (x^\mu x^\nu) - x^\nu \Box x^\mu - x^\mu \Box x^\nu \big) \label{inverseMetric}
\end{equation}

These inverse metric components will define a $d \times d$ matrix at each event in the causal set, which is the inverse metric for the tangent space at that event. To find the metric components at each event, we simply invert each $d \times d$ matrix. This means the metric components are found by requiring $g^{\alpha\beta}g_{\beta\gamma} = \delta^\alpha_\gamma$. 

We will again examine equation \ref{doubleProduct}, this time replacing only $h$ with a coordinate function. This results in 
\begin{align}
    g^{\alpha\beta} (\partial_\alpha f)(\partial_\beta x^\mu) &= \frac{1}{2} \big(\Box (x^\mu f) - x^\mu \Box f - f \Box x^\mu \big) \nonumber\\
    g^{\alpha\mu} \partial_\alpha f &= \frac{1}{2} \big(\Box (x^\mu f) - x^\mu \Box f - f \Box x^\mu \big) \nonumber\\
    d^\mu f &= \frac{1}{2} \big(\Box (x^\mu f) - x^\mu \Box f - f \Box x^\mu \big) \label{dOpDef}
\end{align}
In the last line, we have defined $d^\mu f = g^{\alpha\mu} \partial_\alpha f$. Note that since we can represent $\Box$ and the coordinate functions, we can calculate $d^\mu f$ for any scalar function $f$. 

Finally, note that we can use the metric to relate $d^\mu f$ back to $\partial_\nu f$. 
\begin{align}
    g^{\alpha\mu}\partial_\alpha f &= d^\mu f \nonumber\\
    g_{\mu\nu}g^{\alpha\mu}\partial_\alpha f &= g_{\mu\nu} d^\mu f \nonumber\\
    \delta_\nu^\alpha \partial_\alpha f &= g_{\mu\nu} d^\mu f \nonumber\\
    \partial_\nu f &= g_{\mu\nu} d^\mu f \label{parDerDef}
\end{align}
This allows us to calculate the partial derivatives of any scalar function. Now let us confirm the partial derivatives maintain the assumed property that $\partial_\alpha x^\nu = \delta_\alpha^\nu$. Note by comparing equations \ref{dOpDef} and \ref{inverseMetric}, that $d^\mu x^\nu = g^{\mu\nu}$. Then we have
\begin{align}
    &\partial_\alpha x^\nu = g_{\alpha \mu} d^\mu x^\nu \nonumber\\
    &\partial_\alpha x^\nu = g_{\alpha \mu} g^{\mu\nu} \nonumber\\
    &\partial_\alpha x^\nu = \delta_\alpha^\nu \label{parDeltaProp}
\end{align}

Another property worth checking these partial derivatives for is the product rule. Consider $\partial_\alpha f^2$ for some function $f$. If the product rule held, we would have
\begin{equation}
    \partial_\alpha f^2 = 2f \partial_\alpha f \nonumber
\end{equation}
Now consider a function $d_j$, defined by $d_j(e_i) = \delta_{ij}$. Note $d_j{}^2 = d_j$, so by the product rule
\begin{equation}
    \partial_\alpha d_j = 2d_j \partial_\alpha d_j \label{prodRuleDelta}
\end{equation}
This is only possible if $\partial_\alpha d_j = 0$ everywhere. This alone may not seem like an issue, but any function $f$ can be written as a sum of these $d_j$ functions. 
\begin{equation}
    f = \sum_i f(e_i) d_i \label{deltaSumForF}
\end{equation}
Thus, since the partial derivatives we have defined are linear operators, the only way to satisfy the product rule would be if $\partial_\alpha f = 0$ everywhere for all $f$. Therefore, the partial derivative operators on the causal set will not satisfy the product rule.

Since $\partial_\alpha$ is a linear operator for each $\alpha$, we should be able to represent it as a $n \times n$ matrix. There are two types of linear operators that are applied to a function $f$ in order to calculate the partial derivatives. First, there are matrix operations such as applying the $\Box$ operator. Second, there are operations where you multiply $f$ by a vector component-wise, such as the $x^\alpha f$ term in equation \ref{dOpDef} or multiplying by the metric components in equation \ref{parDerDef}. In order to calculate matrix representations for the $\partial_\alpha$ operator, we will need to represent this second type of operation as a matrix as well. 

Here we will define the map $D : \mathbb{R}^n \to \mathbb{R}^{n \times n}$, that sends a vector to the diagonal matrix with $D(v)_{ii} = v_i$. Note that the component-wise multiplication $v f$ is equivalent to the matrix multiplication $D(v) f$. Now we can convert equation \ref{dOpDef} to a matrix definition of $d^\mu$.
\begin{equation}
    d^\mu = \Box D(x^\mu) - D(x^\mu) \Box - D(\Box x^\mu) \label{dMatrixDef}
\end{equation}
Now using equation \ref{parDerDef}, we can find how to calculate $\partial_\nu$ as a matrix. 
\begin{equation}
    \partial_\nu = \sum_\mu D(g_{\mu\nu}) d^\mu \label{parMatrixDef}
\end{equation}

\subsection{Numerical Results \label{sec:boxNumRes}}

\subsubsection{Partial Derivatives}
To verify the validity of these partial derivatives, we will use numerical tests. First, we choose events by sprinkling into a causal diamond in $\mathbb{M}^2$ at a density $\rho$. The coordinate functions and the $\Box$ operator are then defined using the methods described in section \ref{sec:functionsAndOps} and partial derivative operators are defined as in section \ref{sec:boxDersTheory}.

\begin{figure}[h]
    \centering
    \includegraphics[width=8.6cm]{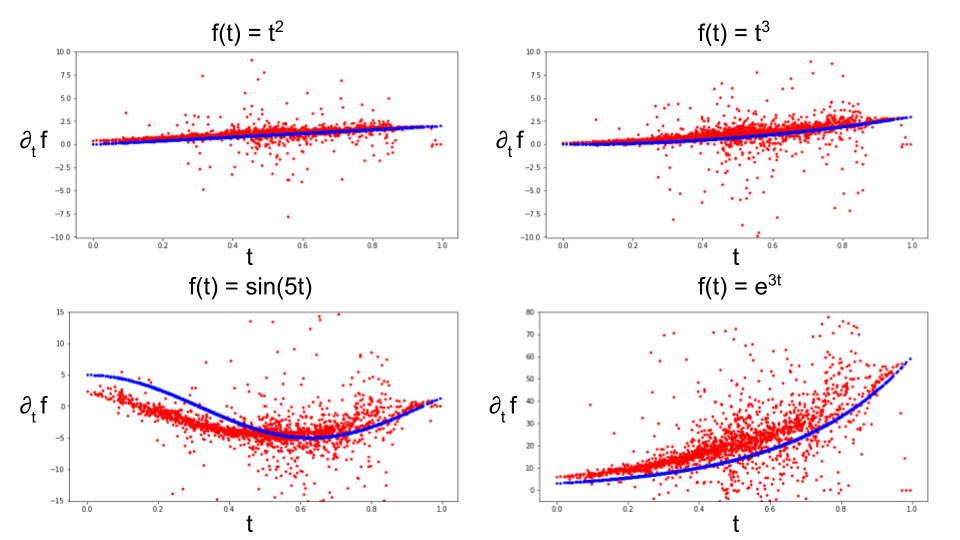}
    \caption{Numerical results for partial derivatives with respect to $t$ (red) and the expected results (blue) for a variety of test functions at a density of $\rho = 4000$.}
    \label{fig:tPlots}
\end{figure}

Figure \ref{fig:tPlots} shows the numerical results for the partial derivatives of a variety of functions of $t$. Clearly, the general shape of the plots match the expected graphs of the derivatives, though there is still considerable error. Note that the error seems most extreme at high values of $t$. This may be happening because these derivatives are defined using the a forward-looking $\Box$ operator. Since events at high values of $t$ have less future events included in $\Box$ calculations, the partial derivatives should be expected to be less accurate. 

\begin{figure}[h]
    \centering
    \includegraphics[width=8.6cm]{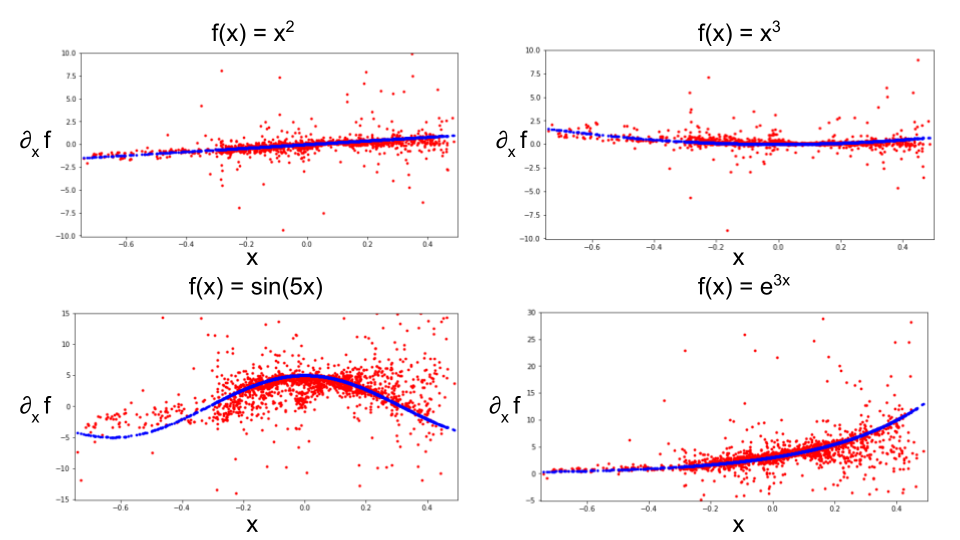}
    \caption{Numerical results for partial derivatives with respect to $x$ (red) and the expected results (blue) for a variety of test functions at a density of $\rho = 4000$.}
    \label{fig:xPlots}
\end{figure}

Similarly, we can test the accuracy of the partial derivatives with respect to $x$. Figure \ref{fig:xPlots}, shows the numerical results for the partial derivatives of a variety of functions of $x$. Again, we see the general shape of the plots matches the expected graphs for the partial derivatives. It also seems that the $x$ derivative graphs have greater error than the $t$ derivatives. This may be because the $\Box$ operator only includes future events, so changes along the $x$ direction may be harder to estimate. 

\begin{figure}[h]
    \centering
    \includegraphics[width = 8.6cm]{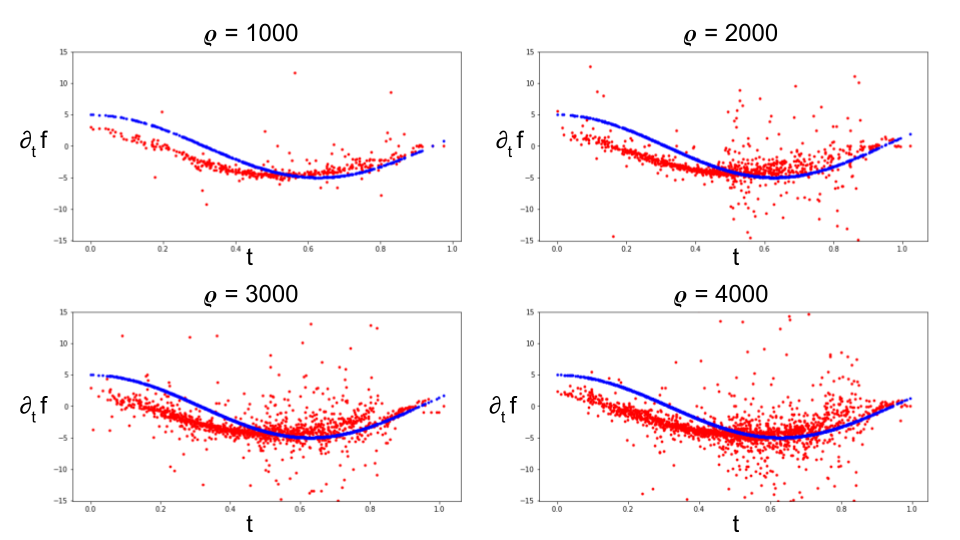}
    \caption{Numerical results for $\partial_t f$ as a function of $t$, with $f(t) = \sin (5t)$, calculated at various densities, $\rho$. The numerical results are shown in red with the expected results in blue.}
    \label{fig:rhoPlots}
\end{figure}

An important question after seeing the error in the derivatives in figures \ref{fig:tPlots} and \ref{fig:xPlots}, is whether this error will decrease at high densities. In figure \ref{fig:rhoPlots}, we see that the error in the derivative of $f(t) = \sin (5t)$ does not change much with the density. Despite this, \cite{SorkinLocality} and \cite{JohnstonEmbedding} suggest that the errors in both $\Box$ and $x^\alpha$ go to 0 at high density. Since these are the only things needed to calculate these derivative operators, it is then reasonable to assume the errors seen here should go away as $\rho \to \infty$. 

\subsubsection{Proper Times}
To test the accuracy of the metric components, we can use them to estimate the proper times. For nearby events, we can relate the proper times to the metric by
\begin{equation}
    \tau^2 = -g_{\alpha\beta}\Delta x^\alpha \Delta x^\beta \label{properTimeCST}
\end{equation}

\begin{figure}[h]
    \centering
    \includegraphics[width = 8.6cm]{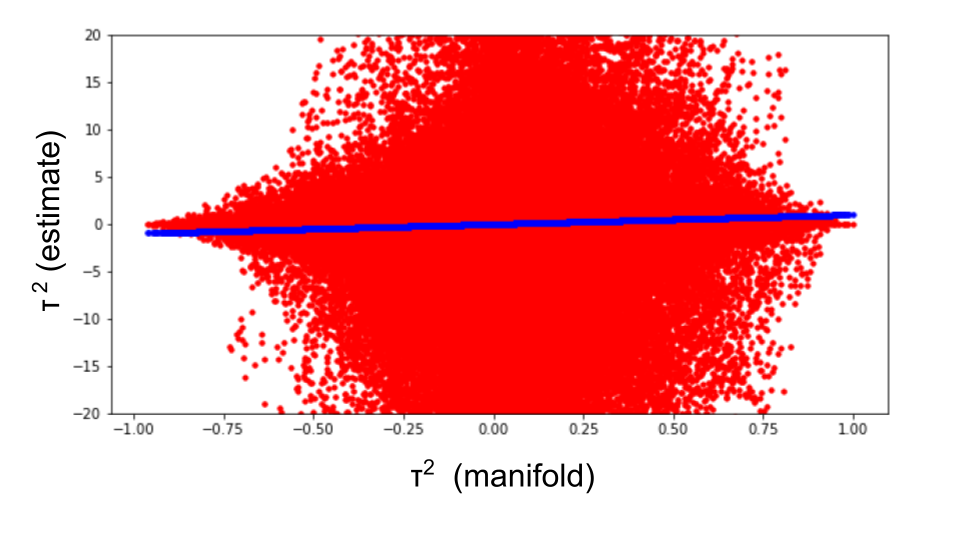}
    \caption{Numerical results for $\tau^2$ (red) plotted against the manifold value of $\tau^2$ (blue). This calculation was done at a density of $\rho = 4000$.}
    \label{fig:tauPlot}
\end{figure}

This can be tested numerically by sprinkling into a causal diamond in $\mathbb{M}^2$ at a density of $\rho = 4000$. The proper time between events is estimated using equation \ref{properTimeCST}, and also calculated using the coordinates of the manifold. The results are compared in figure \ref{fig:tauPlot}, where there appears to be almost no correlation between the manifold proper time and the estimate from the metric. 

\section{\label{sec:geoDers}Partial Derivatives from the Coordinate Space}

\subsection{Theory: Partial Derivatives}
In this section, we will provide an alternative definition of partial derivatives using the Moore-Penrose inverse. Consider a standard linearity condition on the partial derivatives, 
\begin{equation}
    \Delta f = \partial_\alpha f \Delta x^\alpha \label{linCondition}
\end{equation}
We can define the partial derivatives of $f$ at an event $e_i$ in the causal set as the values that most closely match this condition for events near $e_i$. To do this, we must first express this condition as a matrix equation. 

As previously discussed, functions on a region of a causal set can be represented by vectors. If the region has $n$ events then $f$ and each $x^\alpha$ can be expressed as vectors in $\mathbb{R}^n$. We want to find a matrix representation of $\Delta f = f - f_i$, where $f_i = f(e_i)$. This can be accomplished with a matrix,
\begin{equation}
    D_i = \mathbb{I} - \delta_i \label{DMatrix}
\end{equation}
$\delta_i$ in the above definition is the $n \times n$ matrix that is zero except for the $i^{\text{th}}$ column, which is all ones. $D_i$ defined above then satisfies $D_i f = f - f_i = \Delta f$. Similarly, if we define a $d \times n$ matrix $X$ which has the coordinate functions as columns, then $D_i X$ will be a $d \times n$ matrix which has $\Delta x^\alpha$ as columns. 

Using this construction, we can represent equation \ref{linCondition} in matrix form as
\begin{equation}
    D_i f = D_i X 
    \begin{bmatrix}
        \partial_0 f \\
        \vdots \\
        \partial_{d-1} f
    \end{bmatrix}
\end{equation}
Finally, we can solve for the least-squares solution to this matrix equation using the Moore-Penrose inverse. 
\begin{equation}
    \begin{bmatrix}
        \partial_0 f \\
        \vdots \\
        \partial_{d-1} f
    \end{bmatrix} = (D_i X)^{+}D_i f \label{partialFGeoDef}
\end{equation}
Since this is a linear operator on $f$, we can define the partial derivative operators at $e_i$ as the rows of the matrix $(D_i X)^+ D_i$. Note that this definition essentially determines the partial derivatives by enforcing the linearity condition of equation \ref{linCondition}.

The last thing to consider before taking this as our new definition of partial derivatives is how to define the region over which we will apply this linearization. The approximation represented by equation \ref{linCondition} best holds over small regions around the event $e_i$, so it is reasonable to assume our partial derivatives will be most accurate when equation \ref{partialFGeoDef} is applied over a local region. A covariant approach to defining these local regions would be to base them off of the causal structure, perhaps including only links to $e_i$. The approach we apply here uses the coordinate functions more directly. We will include any events which are within a small ball in the coordinate space, centered at $e_i$. More explicitly, events are included in the calculation of equation \ref{partialFGeoDef} only if they satisfy $\sum_{\alpha} (\Delta x^\alpha)^2 < \varepsilon^2$. 

As we will see in section \ref{sec:geoNumRes}, the partial derivatives defined in this way are significantly more accurate than those from the previous section.

\subsection{Theory: Metric Components}
Since we did not generate the metric as a byproduct of finding the partial derivatives like we did in section \ref{sec:boxDersTheory}, we will need to find another way to determine the metric components. Following the theme of the last section, we will try to generate the metric by enforcing a geometric constraint using the Moore-Penrose inverse. In this case, we will force the metric to generate a good estimate of the proper time. 

In some local region near an event $e_i$, the proper time and the metric should be approximately related by the equation 
\begin{equation}
    \tau^2 = -g_{\alpha\beta}\Delta x^\alpha \Delta x^\beta \label{metricCondition}
\end{equation}
To estimate the metric, we will turn this into a matrix equation, where the $\Delta x^\alpha \Delta x^\beta$ terms come directly from the coordinate functions and the $\tau^2$ values are estimated from the causal structure as in \cite{CausalDiamond}. The simplest approach would be to have a $(n \times d^2)$ matrix $S$, with each $\Delta x^\alpha \Delta x^\beta$ making up its columns and a $n$ vector $T$ of the squares of the proper times between each event and $e_i$.
\begin{equation}
    T = -S g
\end{equation}
Here $g$ is the $d^2$ vector of all of the metric components. Then we could find $g$ by using the Moore-Penrose inverse $g = -S^+ T$. This approach nearly works, but there are two issues to correct. 

First, the metric defined in this way will not necessarily be symmetric. To enforce the symmetry of the metric, we can rewrite the $S$ matrix and the $g$ vector to only include the independent components. For example, in two dimensions, we can write this as
\begin{equation}
    \begin{bmatrix}
        \tau_1{}^2\\
        \tau_2{}^2\\
        \vdots\\
        \tau_n{}^2
    \end{bmatrix} = -
    \begin{bmatrix}
        \Delta x^0{}_1 \Delta x^0{}_1 & 2\Delta x^0{}_1 \Delta x^1{}_1 & \Delta x^1{}_1 \Delta x^1{}_1\\
        \Delta x^0{}_2 \Delta x^0{}_2 & 2\Delta x^0{}_2 \Delta x^1{}_2 & \Delta x^1{}_2 \Delta x^1{}_2\\
        \vdots & \vdots & \vdots\\
        \Delta x^0{}_n \Delta x^0{}_n & 2\Delta x^0{}_n \Delta x^1{}_n & \Delta x^1{}_n \Delta x^1{}_n
    \end{bmatrix}
    \begin{bmatrix}
        g_{00}\\
        g_{01}\\
        g_{11}
    \end{bmatrix} \label{metricMatrixCalc}
\end{equation}
Using this modified $S$ matrix and $g$ vector, the metric will be guaranteed to be symmetric. 

The other issue we must address before using this approach to calculating the metric is that we must choose what local region to include in the calculation. We cannot use local coordinate balls like we did with the partial derivatives, because the most accurate estimates for proper time in a causal set can only be applied in the lightcone. Therefore, we will apply this formula only in the overlap of small coordinate balls with the lightcone. This ensures the points are nearby in coordinate space, so that equation \ref{metricCondition} is a good approximation, and that they are within the lightcone, so that we have a good estimate of proper times from the causal structure. 

After fixing these issues, we have now established a method for calculating the metric in a way that is tied to the geometry. 
\begin{equation}
    g = - S^+ T
\end{equation}
Then the inverse metric can be found by inverting the metric $g^{\alpha \beta} g_{\beta \gamma} = \delta^\alpha_\gamma$

\subsection{Numerical Results \label{sec:geoNumRes}}
In this section, we will run numerical tests to assess the validity this approach. We will sprinkle events into a causal diamond in $\mathbb{M}^2$ at a density $\rho$. Then we will calculate the partial derivatives and the metric using the procedures described in the previous two sections.

There is a free parameter that we will need to choose in order to define these derivatives. We must choose values for $\varepsilon$, the size of the balls in the coordinate space used in the calculation of the derivatives and the metric. If $\varepsilon$ is too large, we will effectively be linearizing over the whole interval, so the partial derivatives and the metric will be close to constant. If $\varepsilon$ is too small, the partial derivatives and the metric will be too variable. In the results that follow, we will use $\varepsilon = 0.18$, which I have found to be quite accurate. 

\subsubsection{Partial Derivatives}
\begin{figure}[h]
    \centering
    \includegraphics[width=8.6cm]{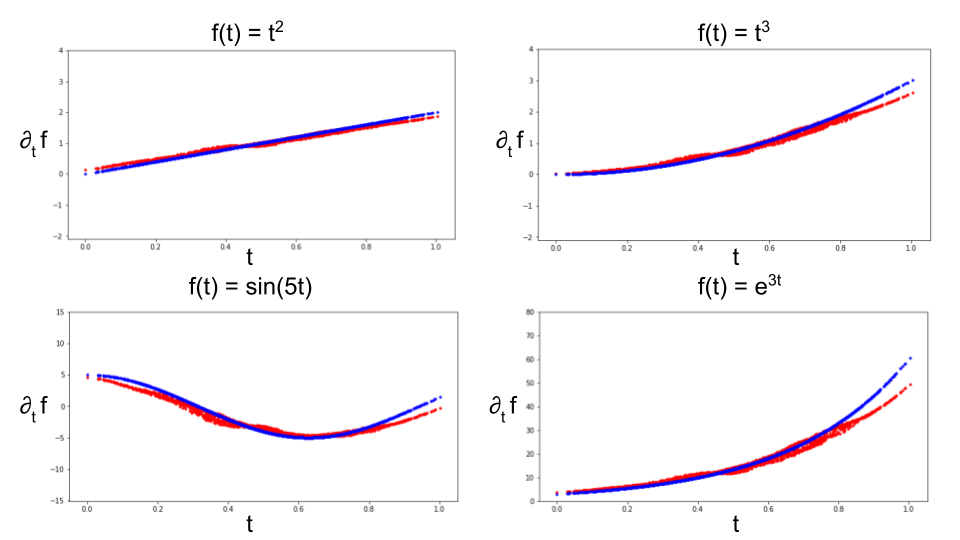}
    \caption{Numerical results for partial derivatives with respect to $t$ (red) and the expected results (blue) for a variety of test functions at a density of $\rho = 4000$.}
    \label{fig:geoTPlots}
\end{figure}

First, we will test the results for the partial derivatives defined in section 5.2. In figure \ref{fig:geoTPlots}, we see the results for $\partial_t f$ for a variety of test functions. These results show close alignment with the expected values, especially when compared to the results found in figure \ref{fig:tPlots} which used the partial derivatives estimated based on the d'Alambertian.

\begin{figure}[h]
    \centering
    \includegraphics[width=8.6cm]{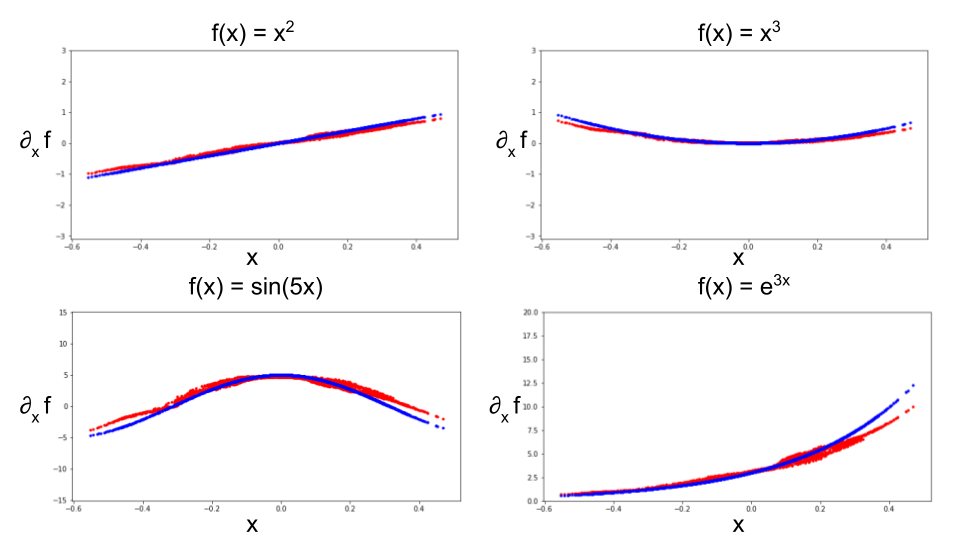}
    \caption{Numerical results for partial derivatives with respect to $x$ (red) and the expected results (blue) for a variety of test functions at a density of $\rho = 4000$.}
    \label{fig:geoXPlots}
\end{figure}

In figure \ref{fig:geoXPlots}, we see the results for $\partial_x f$ for the same test functions. Unlike in figure \ref{fig:xPlots}, the $x$ derivatives with this method are just as accurate as the $t$ derivatives. For both $\partial_x$ and $\partial_t$, we see the greatest errors at the edge of the interval. This is to be expected, since the partial derivatives are calculated by looking at events in a small coordinate ball around each event. When calculating the derivatives near the edge of the interval, there may only be points on one side of the event, which biases the derivative.

\subsubsection{Proper Times}
To test the validity of the metric, we can use it to estimate proper time, as we did in section \ref{sec:boxNumRes}. Again, we see very close alignment with the manifold value, especially when compared to figure \ref{fig:tauPlot}. To some extent, this agreement should be expected, since the metric was calculated to best fit the expected proper times. Note, however, that the only events included in that best fit calculation were timelike separations in a small coordinate ball. Therefore, the correlations seen here for spacelike separations (negative $\tau^2$) and large timelike separations (large positive $\tau^2$) are evidence of the accuracy of this approach to estimating the metric.  

\begin{figure}[h]
    \centering
    \includegraphics[width=8.6cm]{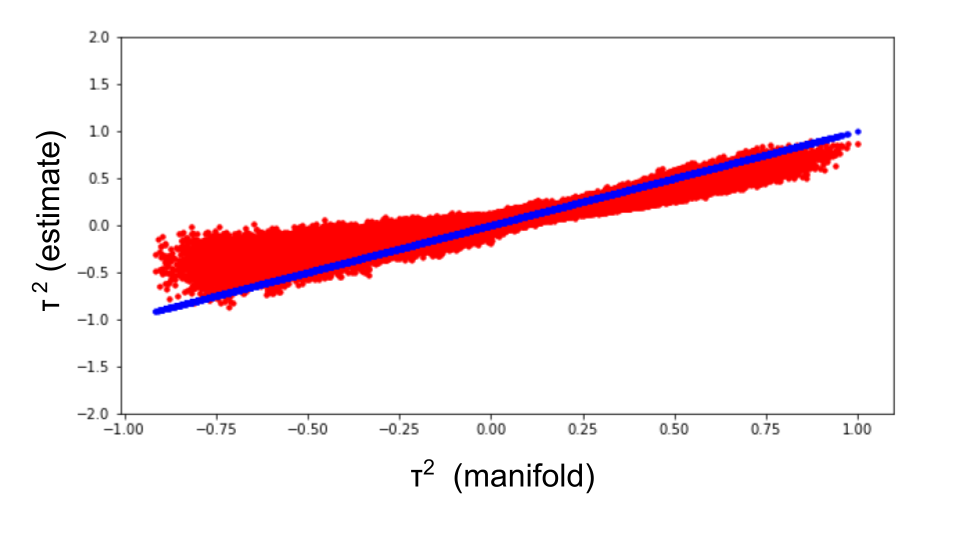}
    \caption{Numerical results for $\tau^2$ (red) plotted against the manifold value of $\tau^2$ (blue). This calculation was done at a density of $\rho = 4000$.}
    \label{fig:geoTauPlot}
\end{figure}

\section{\label{sec:conc}Conclusions}
In this paper, we have devised two methods for calculating partial derivatives on causal sets. The first method involved using the causal set $\Box$ operator defined in \cite{SorkinLocality} in combination with the $\mathbb{M}^2$ coordinate estimators defined in \cite{JohnstonEmbedding}. The partial derivatives found with this method showed reasonable agreement with the expected results for a variety of test functions, but the errors did not seem to decrease as the density increased. This method also produced estimators for the metric components, which were tested by using them to calculate proper times between events. These proper times estimates showed no clear correlation with the manifold proper times from the sprinkling.

The second method for calculating partial derivatives involved using a construction from linear algebra called the Moore-Penrose inverse to find the operators that best fit a linearity condition in small coordinate balls around each event. The results for this method were significantly better than what was found with the $\Box$ operator. We also described another method for defining metric component estimators by using the Moore-Penrose inverse to best fit the proper times in small coordinate balls. This method showed good agreement with the manifold proper time, even for events far outside the small coordinate balls this was optimized on. 

Partial derivatives and metric components are an important part of defining tangent spaces, so this work opens the door to a new framework for estimating geometric quantities. Furthermore, partial derivatives are a key part of many physical theories, such as quantum field theory, and may be useful in that context as well.

\newpage

\bibliography{partialDers}

\end{document}